\documentclass[aps,prl]{revtex4}

\usepackage{epsfig}
\begin{document}
\title{Quantized biopolymer translocation through nanopores: departure from simple scaling}

\author{
Simone Melchionna$^{1,2}$,
Massimo Bernaschi$^3$, 
Maria Fyta$^1$
\footnote{Present address: Department of Physics, Technical University of Munich, Germany},
Efthimios Kaxiras$^{1,4}$, and 
Sauro Succi$^{,4}$
}

\affiliation{
$^1$ Department of Physics and School of Engineering and Applied
Sciences, Harvard University, 17 Oxford Street,
Cambridge, MA 02138, USA\\
$^2$ INFM-SOFT, Department of Physics, Universit\`a di Roma 
``La Sapienza'', P.le A. Moro 2, 00185 Rome, Italy \\
$^3$ Istituto Applicazioni Calcolo, CNR, 
Viale del Policlinico 137, 00161, Roma, Italy \\
$^4$ Initiative in Innovative Computing, Harvard University,
Cambridge, MA, USA\\
}

\date{\today}

\begin{abstract}
We discuss multiscale simulations of long biopolymer translocation through
wide nanopores that can accommodate multiple polymer strands.
The simulations provide clear evidence of folding quantization, namely, the translocation
proceeds through multi-folded configurations characterized by a
well-defined integer number of folds. 
As a consequence, the translocation time acquires a dependence on
the average folding number, which results in a deviation 
from the single-exponent power-law characterizing 
single-file translocation through narrow pores.
The mechanism of folding quantization allows polymers above
a threshold length (approximately $1,000$ persistence lengths
for double-stranded DNA) to exhibit cooperative behavior and as a result 
to translocate noticeably faster.
\end{abstract}

\maketitle


The translocation of biopolymers through nanopores is drawing increasing attention
because of its role in many fundamental biological processes, such as
viral infection by phages, inter-bacterial DNA transduction
or gene therapy \cite{TRANSL}. This problem has motivated a number 
of {\it in vitro} experimental studies, aimed at exploring the 
translocation process through protein channels across cellular
membranes \cite{EXPRM1,EXPRM2}, or through micro-fabricated channels \cite{DEK}.
Recent experimental work has addressed
the possibility of ultra-fast DNA-sequencing using electronic identification of 
DNA bases, while tracking its motion through nanopores under 
the effect of a localized electric field \cite{Branton-Rev}.
Experiments also reported that the translocation 
of biopolymers through pores wide enough to accommodate multiple
strands, exhibits the intriguing 
phenomenon of current-blockade quantization, that is,
discrete jumps of the electric current through the pore
during the translocation process \cite{Li_NatMat2003}.  
This was interpreted as indirect evidence that the 
polymer crosses the pore in the form of discrete configurations, 
associated with integer values of the {\em folding number}, that is, 
the number of strands simultaneously occupying the pore during the translocation.
This behavior has been recently confirmed by direct
observation of multi-folded configurations in large-scale
simulations of biopolymer translocation \cite{ourNL}.

In the present work, we report on the behavior of 
long polymers undergoing translocation through relatively wide pores,
which exhibits qualitatively new features.  Under these conditions, 
we predict from our simulations that folding quantization 
is actually {\it enhanced} and leads to faster translocation 
by effectively reducing friction through 
{\bf the pore region}.
The observed behavior also elicits 
an intriguing analogy with quantum systems, whereby the observed translocation
time can be formulated as a weighted average over the
whole set of multi-folded configurations (the ``pure states'' of the 
polymer-pore system).
Within this picture, the translocation time acquires an additional
dependence on the polymer length, through the average value of the
folding number $\langle  q\rangle_N$, which increases with the polymer length $N$.
Thus, at variance with the case of narrow and short pores \cite{Nelson,Storm_NanoLett2005,ourLBM,Forrey07,PRE08},
translocation through wide pores allowing for multiple simultaneous strands
is not described by a single power-law exponent. 
To give a specific example of scales involved, the size of the pores required 
to observe this behavior in double-stranded  
DNA is of the order of several ($\sim 10$) times the {\em effective}
cross-sectional diameter (which depends on salt concentration and
repulsion between pairs of aligned DNA molecules), while the length threshold 
above which this behavior emerges is $\sim 150,000$ base pairs (bp's).
What is remarkable and counter-intuitive about this behavior is, first, 
{\bf a highly ordered organization of the multiple strands at high folding number},
and second, 
the ability of the quantized configurations to flow through the wide pore without 
experiencing any additional drag, compared to the single-file configuration. 

{\em Multiscale model:}
Our results are obtained from a multiscale treatment 
of translocation, involving a coarse-grained model 
for the biopolymer in which the basic 
unit (``bead'') is equivalent to one persistence length ($\sim 50$ nm), 
and the molecular motion is coupled to the
motion of the solvent in which the biopolymer exists.
Incidentally, the length of polymers considered here, up to 8,000 beads
(equivalent to $\sim 1.2 \times 10^6$ DNA bp's),
is an order of magnitude above any previous simulation in the field.

The model couples microscopic molecular dynamics (MD)
for the biopolymer bead motion to a mesoscopic lattice Boltzmann (LB) \cite{LBE} 
treatment of the solvent degrees of freedom \cite{ourLBM}.
In contrast to Brownian dynamics, the LB approach 
handles the fluid-mediated solvent-solvent 
interactions through an effective representation of local
collisions between the solvent and solute molecules.
The biopolymer translocates through a nanopore under the effect of a 
strong localized electric field applied across the pore ends, similar 
to the conditions in experimental settings \cite{Storm_NanoLett2005},
with the entire process taking place in the {\it fast} translocation regime.

A periodic box of size $N_x  N_y  N_z  (\Delta x)^3$ 
lattice units, with $ \Delta x$ the spacing between lattice points,
contains both the solvent and the polymer.
All parameters are measured in units of the LB
time step and spacing, $ \Delta t$ and  $\Delta x$, respectively (both set to 1);
the MD time step is 0.2.
We take $N_x = N_y = N_z$ with the wall at $x=$($N_x$/2), 
$N_x = 128$ and the number of beads $N$ in the 100 -- 8,000 range. 
At $t=0$ the polymer resides on one side of the separating wall,
$x>$($N_x$/2), near the opening of 
a cylindrical pore of nominal length $l_{p} = 3 $ and nominal diameter $d_{p}$;
we considered a narrower, $d_p=5$, pore and a wider one, $d_p = 9 $.
Translocation is induced by a constant electric field 
acting along the $x$ direction and confined to a cylindrical
channel of the same size as the pore, and length 3 along the 
streamwise ($x$) direction. 
The pulling force associated with the electric field, $E$,
in the experiments is $q_eE = 0.02$ and the average thermal speed  $k_B T/m=10^{-4}$, where
$q_e$ is the effective charge per bead.
The interaction between monomers and with the wall are modeled by 
6-12 Lennard-Jones potentials~\cite{LJ-parameters}, and other aspects of the simulation are
the same as those in our previous work~\cite{ourLBM},
which successfully reproduced single-file translocation~\cite{LBE-parameters,PRE08}.
The effective width and radius of the surrounding pore must take into
account the repulsive bead-wall interactions
{\bf that result in an effective exclusion distance of $\simeq 1.5$  ~\cite{ourLBM}. Therefore,}
a monomer is considered
to lie inside the pore if contained in a pore of effective width $l^{eff}\simeq 6$ 
{\bf (due to exclusion on both sides of the separating wall)}
and 
diameter $d^{eff}\simeq 7.5$ for $d_p=9$ ($d^{eff}\simeq 3.5$ for $d_p=5$).
To measure the residence number of beads in the pore region we define a
cylinder of length $h_{p}=10$ and radius $d_p$ centered at the pore midpoint 
and with axis aligned with the pore. 
This extended region misses monomers close to the
pore openings and in contact with the wall, 
but  permits to measure the number of beads in a wider region than the pore width
with better statistics (reduced variation in the resident-bead number).

\begin{figure}
\includegraphics[width=0.5\textwidth]{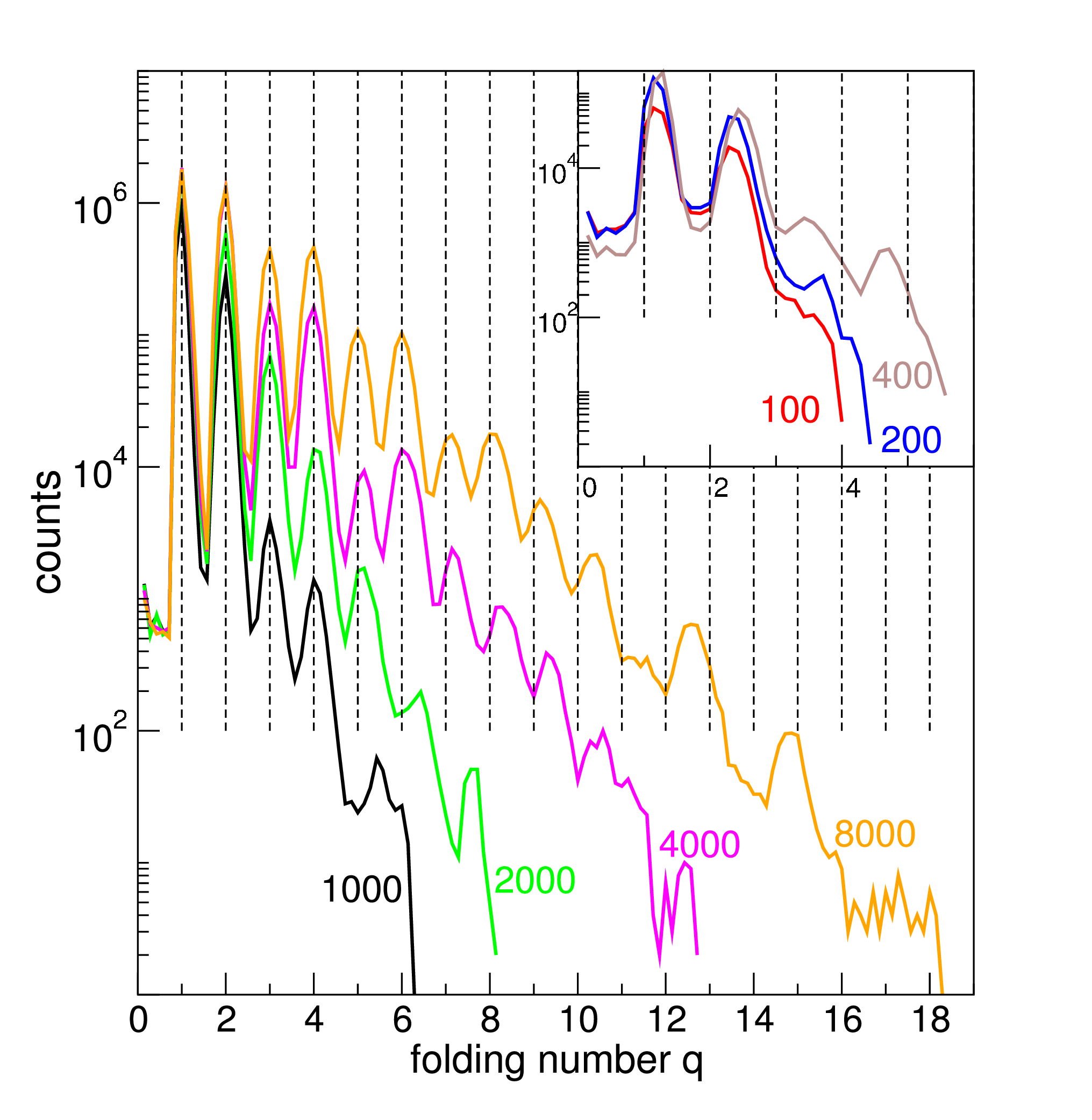}
\caption{(Color) Probability distribution of the folding number $q$ for polymer lengths 
$N=1,000-8,000$ and pore width $d_p=9$. 
Peaks at integer $q$ values (shown by vertical dashed lines) are evident.  The inset 
shows data for $N=100-400$ at low-$q$ values.
}
\label{fig:q_lp9}
\end{figure}

{\em Configurational analysis - quantization of the folding number:}
In Fig.\ref{fig:q_lp9}, we show the cumulative statistics of the 
folding number, $q=N_{res}/N_{1}$, collected at each time-step of every 
single trajectory for a series of $100$ realizations for each polymer length.
Here, $N_{res}$ is the number of resident beads at each given time for each 
realization, while $N_{1}$ is the observed single-file value of the resident number, 
$N_{1}=h_p/b^{eff} \equiv h_p (1/b+1/\sigma_b)/2 \sim 7$ (for values of 
$\sigma_b$ and $b$ see \cite{LJ-parameters} and \cite{LBE-parameters}).
The combined statistics over initial conditions and time evolution, produces 
an aggregate ensemble, ranging from about $10^5$ time frames for the 
shortest history, ($N=100$, $d_p=9$), up to over $10^7$  time frames 
for the longest one ($N=8,000$, $d_p=5$).
The case $d_p=9$ reveals a sharp quantization of the distribution of the
folding number, with well defined peaks, which
closely resembles the one observed in double-strand DNA 
translocation through solid nanopores (see Fig.7 of \cite{StormPRE}).
For the shortest strands, $N=100,~200,~400$, the peaks are shifted
to values slightly larger than the integer values $q=1, 2$. 
This indicates that the polymer spends most of its time between the low-fold states
$q=1, 2$.
However, for longer strands $N>1,000$, the peaks of the distribution
appear almost perfectly centered at integer values, up to $q=4,5,6$, respectively.
This quantized spectrum is particularly evident for the case of
the longest polymers, $N=4,000$ and $8,000$.
In this case, ``quantum states'' up to $q=10$ are populated, with a 
slight shift of the peaks to values higher than integer values only for $q>7$.
Note that the quantization is evident up to $q=10$, which is still significantly smaller than 
the largest folding number compatible with the pore 
diameter, $q_{max} = (d_p/\sigma)^2 \sim 20$ \cite{QMAX}. 
The value of $q$ above which the quantization gradually diverges
from integer values, is an increasing function of the polymer length.
The case $d_p=5$ (not presented) shows the same structure, though on a smaller range
of $q$ values, up to $q=5$.

\begin{figure}
\includegraphics[width=0.45\textwidth]{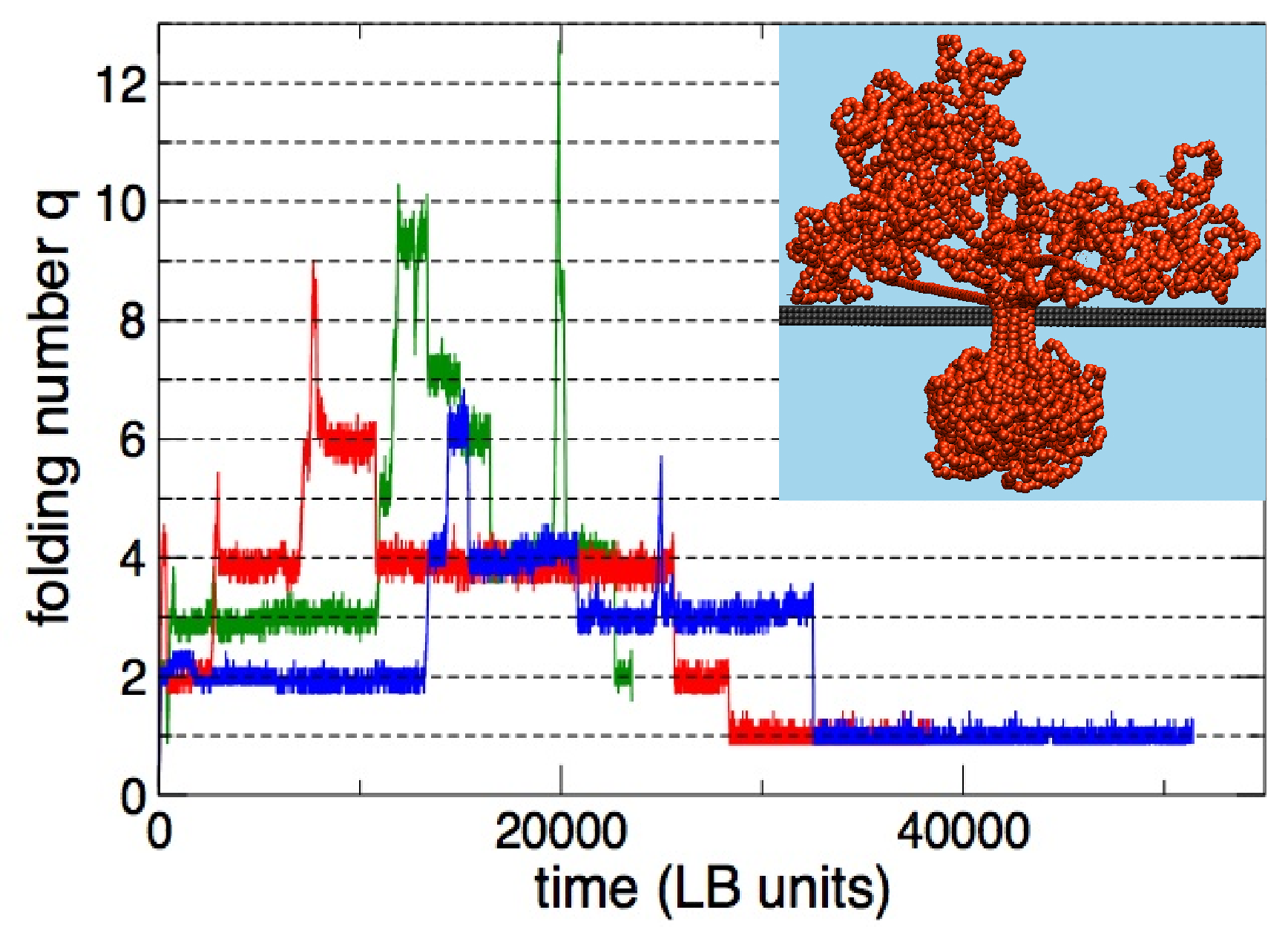}
\caption{(Color) Three trajectories of the folding number sampled from
the ensemble $N=4,000$ and $d_p=9$. 
Low $q$ configurations are most often visited, but $q\geq 10$
can also occasionally occur. 
The inset shows an actual translocation event at the mid-point with $q=9$.
}
\label{fig:nres}
\end{figure}

These data suggest an intriguing analogy with quantum systems, the 
folding number playing the role of the quantum numbers associated 
with excited states of atomic and molecular configurations. 
Within this analogy, single-file translocation ($q=1$) would represent the 
analog of the ground state of the polymer-pore system.
Indeed, the long-term, final stage of the translocation, is always found to
proceed in single-file mode, corresponding to the polymer tail. 
In this respect, long polymers translocating through
wide pores are naturally expected to exhibit a richer spectrum 
of excitations vs. short polymers translocating through narrow pores.
In particular, the number of excited states supported by the polymer-pore system should
grow quadratically with the pore diameter.
Specifically, a polymer of length $L=bN$ translocating through a pore of length $l^{eff}_p$ and
diameter $d^{eff}_p$, can produce a spectrum of folding numbers up to a maximum of
$q_{max} \propto (d^{eff}_p/\sigma_b)^2$ strands (full-packing limit), each consisting of 
$N_{1} = l^{eff}_p/b^{eff}$ monomers.
This limit can only be saturated by sufficiently long polymers, such that 
$N/N_{1} \gg  q_{max}$, namely $N \gg N_{p} \sim l^{eff}_p (d^{eff}_p)^2/b \sigma_b^2$, with
$N_{p} \sim 100$ the saturation length in the present work.

The statistical picture presented above is supported by 
the dynamic trajectories. In Fig. \ref{fig:nres}, 
we show the time-evolution of the folding number $q(t)$ for 
three histories drawn at random from the pool of a hundred realizations of the $N=4,000$ 
system.  A very rich dynamics, with sudden jumps between the various 
``excited states'', is clearly visible. Interestingly, jumps occur both ways, 
from low to high $q$ and vice-versa, 
corresponding to absorption/emission of ``fold-quanta''. 
This indicates that translocation is not monotonic, but consists 
of a mixed sequence of folding and un-folding events.
As anticipated, this sequence is always found to end up in single-file
configuration, corresponding to the translocation of the polymer tail.
Interestingly, the trajectories spend virtually all of their time in 
quantized states with very sharp transitions between them. 
A snapshot of such a quantized state for a highly folded
configurations ($q=9$) is shown in the inset of Fig. \ref{fig:nres}.

\begin{figure}
\includegraphics[width=0.5\textwidth]{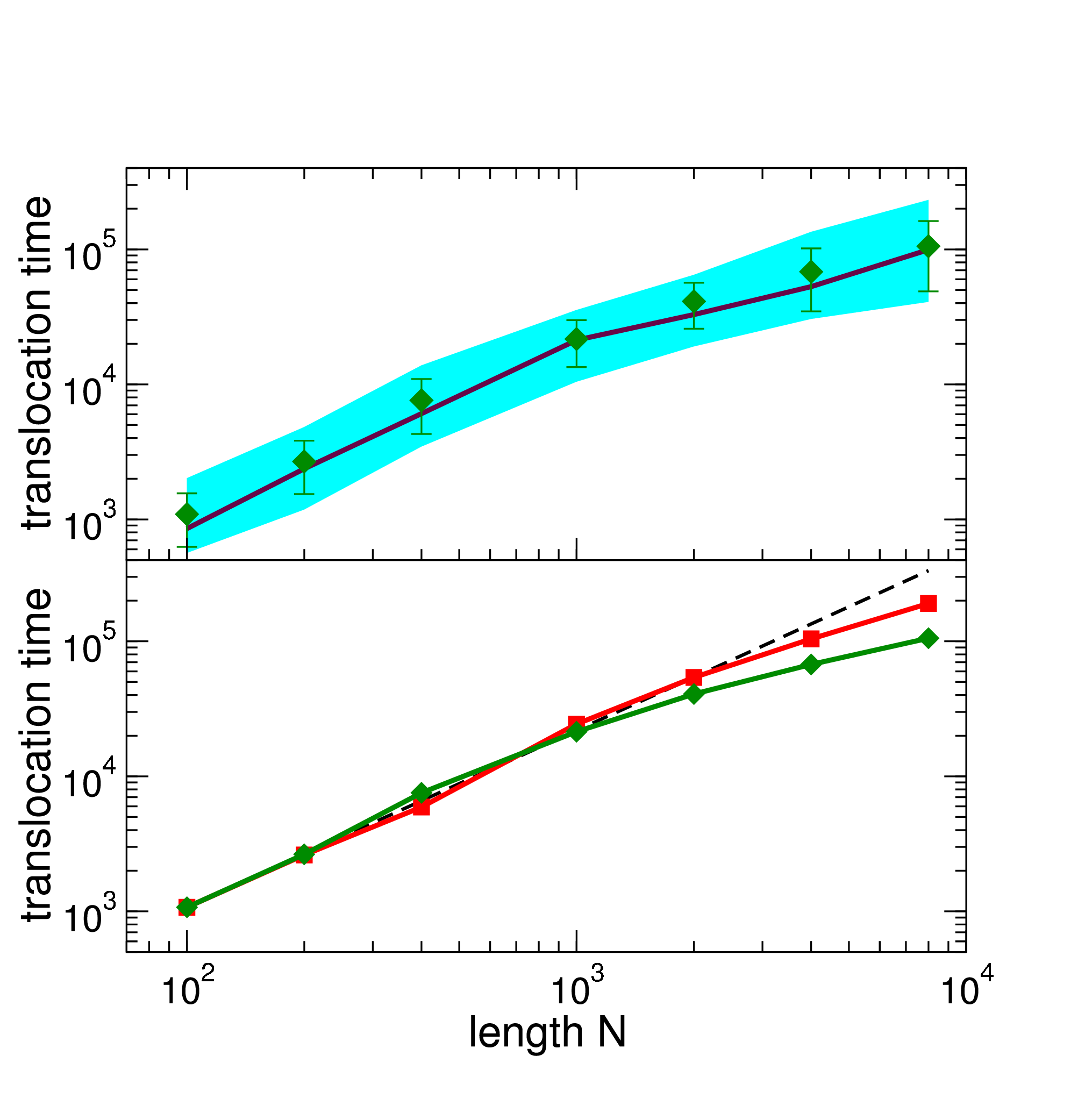}
\caption{(Color) Translocation times  
as a function of the polymer length ($N$) for wide pore ($d_p=9$).
Top panel shows the average translocation time (green diamonds 
with associated error bars), the most probable translocation
time (purple line) and the range between the maximum and minimum
translocation times (shaded region). 
Bottom panel shows a linear 
fit to the low range results $10^2<N<10^3$ (dashed line),
the multifile translocation time $\tau_{mf}(N)$ from 
Eq. (\ref{TAUINT}) (red squares), 
and the average translocation time (green diamonds).
}
\label{fig:Translotime}
\end{figure}

{\em Scaling exponents:}
In Fig. \ref{fig:Translotime}, we report the translocation time as a 
function of the polymer length, $N$, for translocation through the wide pore
($d_p=9$). The most probable time from the time distributions,
together with the average and the range between the minimum and maximum times are presented.
From this figure, it is apparent that up to a length $N=1,000$, the translocation time
$\tau$ obeys a scaling law of the form $\tau \sim N^{\alpha}$, with 
$\alpha \sim 1.36$ for the most-probable translocation time,
slightly larger than the corresponding value for narrow 
pores~\cite{ourLBM,Storm_NanoLett2005,PRE08,Forrey07,Nelson}. 
We emphasize that, regardless of the exact values of the scaling 
exponents, all translocation indicators, that is, 
the minimum, maximum, most probable, and average translocation times, 
point clearly towards a deviation from a single-exponent power-law in the region $N>1,000$,
a clear signature of multi-fold translocation.   
By restricting the analysis to the four longest 
chains, $N=1,000,~2,000,~4,000,~8,000$, 
the bending of the curve (reduced translocation time) 
might be interpreted as the emergence of a new  scaling exponent, $\alpha_2 \sim 0.75$. 
However, as shown below, this bending is due to the multi-fold 
conformation of the translocating biopolymer, which does not necessarily 
follow a power-law dependence on the polymer size.

\begin{figure}
\includegraphics[width=0.5\textwidth]{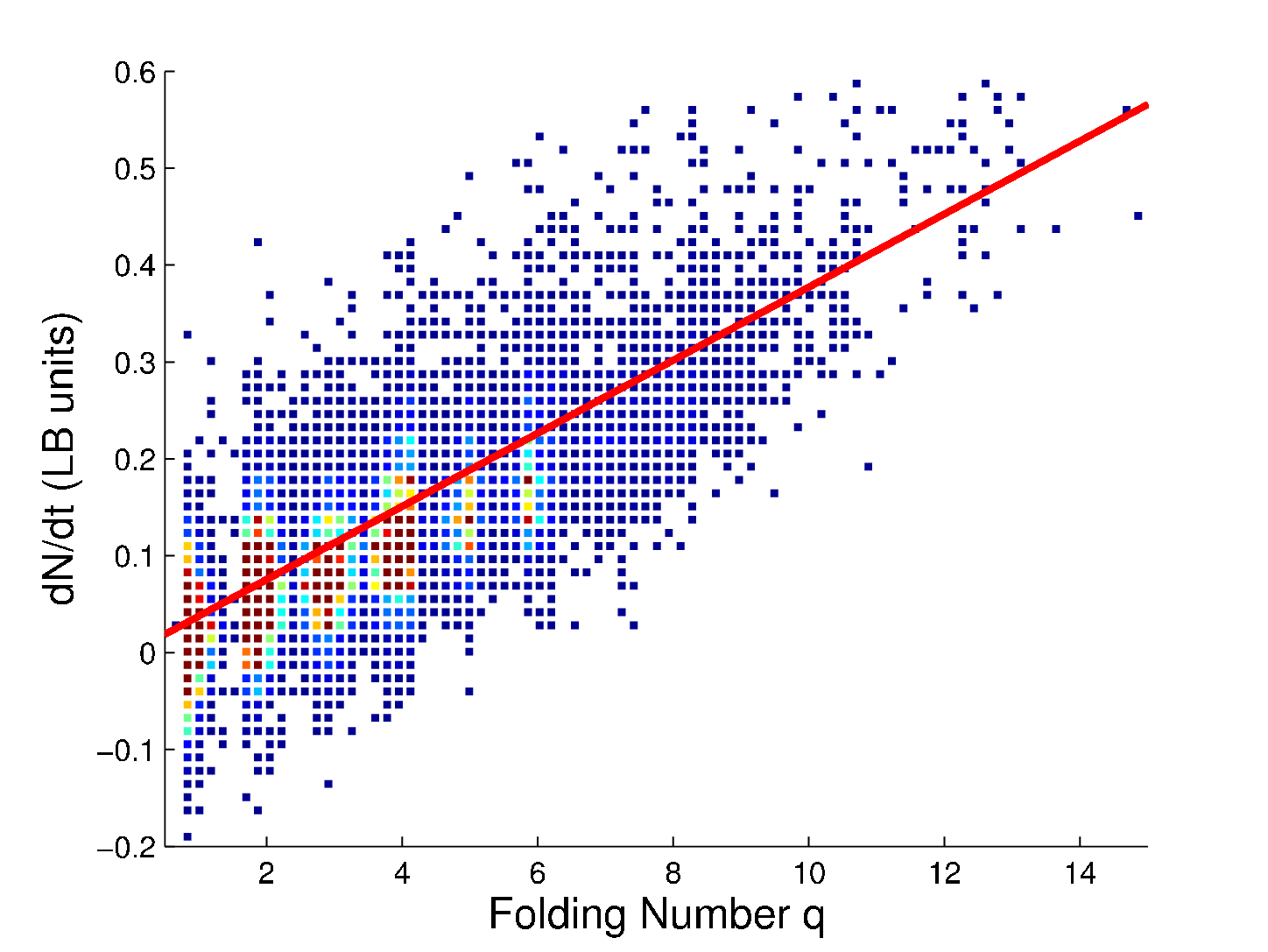}
\caption{
(Color) Scatter plot illustrating the correlation between the folding number $q$ and
the rate of translocating monomers $dN/dt$ for the wide pore ($d_p=9$) and
long polymer ($N=8,000$) case. 
Colors indicate the density of points (blue: low, red: high);
the straight line is the best linear fit.
}
\label{fig:fric}
\end{figure}

The translocation dynamics depends on the strength of the frictional forces exerted by the wall.
In the single-file scenario, strong friction can change the power-law exponent from
$\simeq 1.2$ to a linear relation $\tau \propto N$ \cite{diventra}. 
In the case of multi-file translocation, a central issue is whether or not 
the highly folded configurations induce high-friction conditions.
As demonstrated in Fig.~\ref{fig:fric}, $dN/dt$ is linearly correlated to $q$, 
with approximately the same slope for all folds.  If friction were dominant, 
$dN/dt$ vs. $q$ would asymptotically reach a constant value with increasing $q$, 
which is clearly {\em not} observed in the simulations.
Moreover, the $dN/dt$ vs. $q$ slope depends 
only slightly on the polymer length (data not shown), changing by $\sim 30$\% in going from 
$N=400$ to $N=8,000$, which is further evidence in support of hydrodynamic coherence 
inside the pore.  
{\bf It is likely that such coherence arises from small velocity differences between
neighboring beads in the pore, by minimizing bead-bead frictional forces,
and/or the lubricating effect of the surrounding solvent enhanced by the alignment of strands.}
Therefore, frictional forces have a negligible effect, possibly
limited to a small layer close to the wall, and unimportant for the group of 
translocating monomers. This rules out the possibility that the change of exponent is caused by
frictional forces inside the pore.


Regardless of the underlying nature of the translocation process, we can compute
the translocation time of each realization, with $\tau$ the total translocation time:
\begin{equation}
N = \int_0^{\tau} \frac{dN}{dt} dt = K_N \int_0^{\tau} N_{res}(t) dt
\label{NINT}
\end{equation}
where we have used the fact that $dN/dt$ and $N_{res}$ are linearly 
correlated with a constant of proportionality $K_N$. 
Next, we write $N_{res}(t)=q(t) N_{res,1}$, where the subscript 1 stands 
for the single-file ($q=1$) limit of the residence number. 
Eq. (\ref{NINT}) then leads to $N =  K_N N_{res,1} {\bar q}_N \tau$, where
${\bar q}_N$ is the time averaged value of $q$.
By averaging over all realizations, we obtain:
\begin{equation}
\tau_{mf}(N) \sim  \frac{N}{N_{res,1}\langle K_N \rangle  \langle {\bar q}_N \rangle} 
\equiv \frac{\tau_{1}(N)}{\langle {\bar q}_N \rangle}
\label{TAUINT}
\end{equation}
where brackets stand for ensemble averaging, $\tau_{mf}(N)$ is the multi-file translocation time, and
we have defined $\tau_1(N) \equiv (N/N_{res,1})(1/\langle K_N \rangle) $; note that the
dependence of $\langle K_N \rangle$ on $N$ is responsible for the non-linearity of $\tau_1(N)$.
The simulation data show that the average $\langle {\bar q}_N \rangle$ remains approximately
constant $\sim1.2$, for $N < 1,000$, and then begins growing, reaching $\sim2.6$ for $N=8,000$.
Therefore, we conclude that the departure of the translocation
time from a power-law at large $N$ is mainly due to the increase of 
$\langle {\bar q}_N \rangle$ with polymer length, for $N>1,000$. 
This, in turn, results from the shift of the probability distribution of the translocation time 
towards higher $q$ values as $N$ is increased.
For all lengths $N$ considered here, the time average $\langle {\bar q}_N \rangle$ remains below
$3$, because the states $q=1$ and $q=2$ continue to be the most populated ones, 
for both pore diameters $d_p = 5, \; 9$.
We have also checked that the high-$q$ peaks have a sizeable
effect only on moments $\langle q^p \rangle$ for $p \geq 5$.
Since the average translocation time is only a first order moment, the quantized 
peaks have little effect on it. 
This explains why the two pores, $d_p=5, \; 9$, show similar
dependence of $\langle \bar q_N \rangle$ on $N$.
As a self-consistency check, in Fig. \ref{fig:Translotime} we show the 
average translocation time for the case $d_p=9$, and compare it to the  
single-exponent estimate $\tau_{1}(N) \sim N^{1.31}$ 
and the compensated multi-file translocation time, $\tau_{mf} = \tau_{1}(N)/\langle \bar q\rangle_N$.
The reasonable match of $\tau_{mf}$ with the data  supports the idea that the speed-up of the
longest chains can be attributed to the spectral shift of the folding number $q$. 

{\bf 
Finally, we note that analyzing the in-pore conformation vs. $q$
is an important aspect of multi-file translocation. 
In the present communication we focused on how translocation, and the accompanying
single-file out-of-pore hydrodynamics, is modulated by the population of the in-pore states, 
but have not analyzed in detail the in-pore conformations, which will be addressed in 
future work.  
}

\acknowledgments{
SM and MB acknowledge support by Harvard's Initiative in Innovative Computing.
Computational resources were provided through the 
CyberInfrastructure Lab of the Harvard School of Engineering and Applied Sciences.
We thank G. Lakatos for helpful comments.}


\begin{thebibliography}{99}

\bibitem{TRANSL} H. Lodish, D. Baltimore, A. Berk, S. Zipursky, 
P. Matsudaira, and J. Darnell, {\it Molecular Cell Biology}
(W.H. Freeman \& Co, NY, 1996).

\bibitem{EXPRM1} J.J. Kasianowicz {\it et al.}, Proc. Nat. Acad. Sci. (USA), 
{\bf 93}, 13770 (1996).

\bibitem{EXPRM2} A. Meller {\it et al.},  Proc. Nat. Acad. Sci. (USA) {\bf 97}, 1079 (2000).

\bibitem{DEK} C. Dekker, Nature Nanotech., {\bf 2}, 209, 2007.

\bibitem{Branton-Rev}
For a recent review of the field see
D. Branton {\it et al.}, Nature Nanotech., {\bf 26}, 1, 2008.



\bibitem{Li_NatMat2003}
J. Li  {\it et al.}, Nat. Mater. {\bf 2}, 611 (2003).

\bibitem{ourNL} M. Bernaschi, S. Melchionna, S. Succi, M Fyta, and E. Kaxiras,
Nano Lett. {\bf 8}, 1115 (2008).

\bibitem{Nelson}  D.K. Lubensky and D.R. Nelson, Biophys. J. {\bf 77}, 
1824 (1999).

\bibitem{Storm_NanoLett2005}
A.J. Storm {\it et al.},  Nano Lett. {\bf 5}, 1734 (2005).

\bibitem{ourLBM} M.G. Fyta, S. Melchionna, E. Kaxiras, and S. Succi,
Multiscale Model. Simul. {\bf 5}, 1156 (2006).

\bibitem{Forrey07} C. Forrey and M. Muthukumar, J. Chem. Phys. {\bf 127}, 015102 (2007).

\bibitem{PRE08} M. Fyta, S. Melchionna, S. Succi, E. Kaxiras, Phys. Rev. E {\bf 78}, 036704 (2008).

\bibitem{LBE} R. Benzi, S. Succi, and M. Vergassola,
Phys. Rep. {\bf 222}, 145, (1992).

\bibitem{LJ-parameters}
The Lennard-Jones parameters for bead-bead interactions are 
$\sigma_b=1.8 $, $\epsilon_b=10^{-4}$ and cut-off distance $2.02$;
for bead-wall interactions
$\sigma_w=1.5 $, $\epsilon_w=10^{-3}$ and cut-off distance $1.68$.

\bibitem{LBE-parameters}
Briefly, the bonds between adjacent beads are modelled by springs with
constant $k=0.5$ and equilibrium length  $b=1.2$; the solvent density is $\rho=1$,
kinematic viscosity $\nu=0.1$ and friction coefficient $\gamma=0.1$ (all in LB units).
With $\Delta x=42$ nm,  $b$ becomes equal to the
persistence length of double-stranded DNA ($50$ nm). 

\bibitem{StormPRE} A.J. Storm {\it et al.}, Phys. Rev. E {\bf 71}, 051903 (2005).

\bibitem{QMAX} To be precise, $q_{max}= (d_p-\sigma_{w})^2/\sigma_{b}^2 = 17.36$
for $d_p=9$; but beads can get closer to each other than $\sigma_{b}$ and closer
to the wall than $\sigma_{w}$, giving $q_{max} \sim 20$.


\bibitem{diventra}
M. Zwolak and M. Di Ventra,  Rev. Mod. Phys. {\bf 80}, 141 (2008).

\end{thebibliography}
\end{document}